\documentclass[preprint,3p,times,twocolumn]{elsarticle}

\usepackage{amssymb}
\usepackage{amsmath}
\usepackage{mathtools,cuted}
\usepackage[english]{babel}
\usepackage{hyperref}
\usepackage{xcolor}
\usepackage[normalem]{ulem}

\journal{Computer Physics Communications}

\begin{document}

\begin{frontmatter}

\title{Implementation of photon partial distinguishability in a quantum optical circuit simulation}

\author[a,b]{Javier Osca\corref{corr}\fnref{email1}}
\cortext[corr]{Corresponding author.}
\fntext[email1]{javier.oscacotarelo@mu.ie}
\author[a,b]{Jiri Vala\fnref{email2}}
\fntext[email2]{jiri.vala@mu.ie}
\address[a]{Department of Theoretical Physics, Maynooth University, Ireland}
\address[b]{Tyndall National Institute, University College Cork, “Lee Maltings,” Dyke Parade, Cork, Ireland}

\begin{abstract}
We are concerned with numerical simulations of quantum optical circuits under certain realistic conditions, specifically that photon quantum states are not perfectly indistinguishable. The partial photon distinguishability presents a serious limitation in implementation of optical quantum information processing. In order to properly assess its effect on quantum information protocols, accurate numerical simulations, which closely emulate quantum circuit operations, are essential. Our specific objective is to provide a computer implementation of the partial photon distinguishability which is in principle applicable to existing simulation techniques used for ideal quantum circuits and which avoids a need for their significant modification. Our approach is based on the Gram-Schmidt orthonormalization process, which is well suited for our purpose. Photonic quantum states are represented by wavepackets which contain information on their time and frequency distributions. In order to account for the partial photon distinguishability, we expand the number of degrees of freedom associated with the circuit operation extending the definition of the photon channels to incorporate wavepacket degrees of freedom. This strategy allows to define delay operations in the same footing as the linear optical elements. 
\end{abstract}

\end{frontmatter}

\section{Introduction}
Quantum computers are becoming a reality. They are emerging in various platforms due to considerable progress in different fields. These include platforms based on superconducting qubits, trapped ions, or neutral atoms \cite{General,Super,Trapped} to provide just a few examples. One attractive approach \cite{KLM,Linear} relies on photon states to encode qubits on linear optical elements, such as phase shifters and beamsplitters, 
to perform different quantum operations. It is known that linear optical elements alone are not sufficient to represent an arbitrary quantum operation and have to be complemented by post-selection to achieve universality.

Due to physical limitations the first optical quantum circuits are built for  
specific purpose-oriented functionalities. Moreover their operations are approximate due to imperfections and noise processes. Given these circumstances, we consider it important to study the feasibility and behaviour of those circuits under realistic conditions.

Partial distinguishability is one of the most common sources of imperfections in optical circuits. It originates, for example, from imperfect synchronization of photons or a mismatch in their spectra due to imperfections in the photon sources. Our interest is to determine how the output of an ideal circuit is modified if photons are not perfectly indistinguishable. We require the simulation output to be described by a set of probability amplitudes that includes their relative phases which play important roles in operations relevant to quantum computation and communication protocols. For example, the key element of the quantum computation with linear optical elements \cite{KLM} is implementation of the conditional phase flip operation, and more complex protocols such as entanglement swapping or teleportation \cite{Entanglement} rely on specific relative phase relations between probability amplitudes.

There are two kinds of software platforms to simulate quantum optics. One is based on continuous variable models where physical observables, like the strength of an electromagnetic field, are calculated \cite{Strawberry,FPAQS}. The other, which is relevant to this work, is based on the Fock description of photonic states \cite{Perceval,PQLAB,SOQCS}. At the moment of writing this article, none of the publicly available libraries implement a general mechanism to calculate partial distinguishability of photons or general delay gates. Calculations involving partial photon distinguishability appear to be limited to either ad-hoc approaches or to simple cases as examples.

Very efficient methods dealing with the partial photon distinguishability can be found in the context of boson sampling \cite{BS1,BS2,BS3}. However, these methods are focused on sampling (and probability distributions obtained via repeated sampling) where no phase information regarding output quantum states can be recovered. Furthermore, the way in which partial photon distinguishability is considered in these methods is intertwined with the way the samples are obtained.

An important attribute of our approach is its modularity which allows its integration into an existing simulation framework. This also means that additional physical considerations may potentially be done  without the need for its complete reconstruction. Our objective is twofold. First, to implement a modular simulation method for partial photon distinguishability that can be used jointly with libraries capable of calculating the output Fock state of ideal optical quantum circuits and, second, to make this implementation intuitive enough, so that its use is compatible with physical definitions of photon states. 

Our approach is based on extending the channel definition of an ideal circuit to consider different photon wavepackets. The number of wavepackets used is reduced to a finite manageable subset relevant for the circuit simulation and each one of them is labelled by an index. This index is treated in the same way as any other quantum number in the course of a simulation. This strategy allows us to handle partial distinguishability in the emitter and the delay operation on the same footing as any of the linear optical elements. This is the case even if those operations are not unitary, as discussed in detail below.  The partial distinguishability in the emitter is calculated using a Gram-Schmidt orthornormalization \cite{Distin} of the wavepackets. The resulting orthonormalized packets are also an integral part of the delay definition. To obtain automatically the relevant coefficients
from a physical definition of the photon states, software is structured into different layers of abstraction.

The present paper stems from the development of our Stochastic Optical Quantum Circuit Simulator (SOQCS) library \cite{SOQCS} which was written using the C++ programming language. We choose to perform the implementation with our own library, nevertheless the approach, presented here, is general and can be used to automate partial distinguishability calculations with any library which is capable of computing the output state of an ideal optical circuit. Note that the SOQCS library contains various methods to perform these computations. Furthermore, it also contains mechanisms to calculate other effects such as photon losses, detector imperfections and to perform some post-processing on the output. The essential feature of the library is a separation of different effects, such as partial distinguishability in particular, into distinct modules. Description of the library and its structure, and details about the interplay of the distinguishability model with the additional effects are available in the public release of the SOQCS library \cite{SOQCS}. 

\begin{figure}[h]
  \centering
 \includegraphics[width=0.4\textwidth]{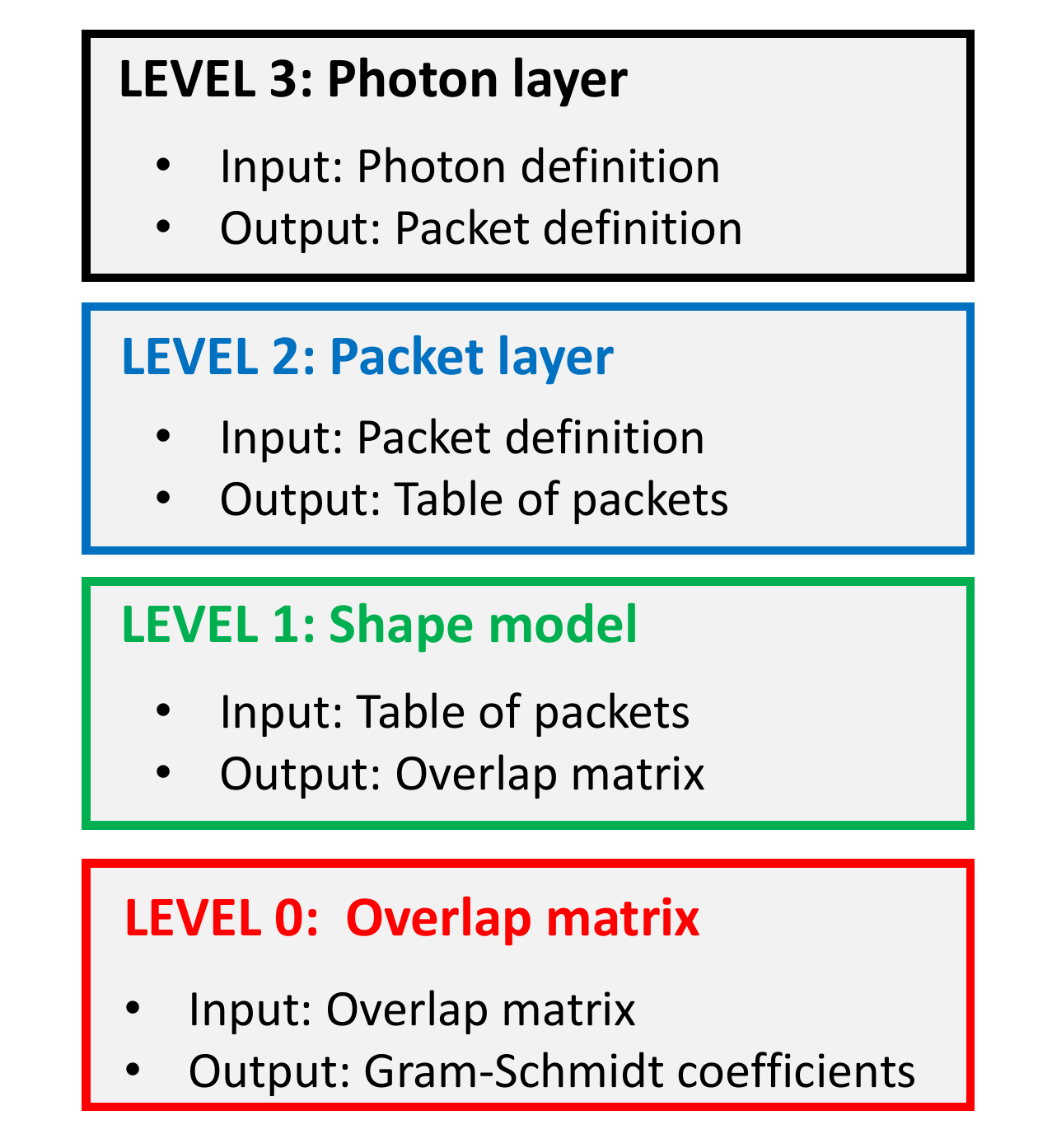}
  \caption{ Levels of abstraction in the implementation of the partial distinguishability between photons in our numerical simulation. Level 0: This level concerns the definition of the distinguishability in terms of wavepacket overlaps. Level 1: A photon wavepacket shape model is used among various to choose and its properties are summarized in a single table of parameters. Level 2: Packets are defined as an abstraction than can be configured individually. Level 3: Photons are created as an abstraction that contains information about the state occupation and the characteristics of its degrees of freedom.} 
  \label{F1}
\end{figure}

The structure of the paper mirrors the layers of abstraction needed to calculate and use the Gram-Schmidt coefficients (bottom layer) from a physical definition of photons and gates (top layer) as illustrated in Fig. \ref{F1}.  In the first two sections of the paper we will review basic concepts, including the definitions of quantum optical circuit, state and simulator (Sec. \ref{One}) and introduce the underlying mathematical formulation of distinguishability (Sec. \ref{Two}). In the next sections, we explain how partial distinguishability on the emitter and delays is calculated using the mathematical formulation (in the form of Gram-Schmidt coefficients) from a more intuitive definition of photons and devices. In Sec. \ref{Three} it is shown how those Gram-Schmidt coefficients are obtained from the overlaps between photon wavepackets. In Sec. \ref{Four} a photon wavepacket shape model is introduced that can be configured using a specific table of parameters. In Sec. \ref{Five}, we explain how this information can be represented as a photon definition in a virtual optical circuit. We show some simple examples using the SOQCS library \cite{SOQCS} in Sec. \ref{Six} before concluding in Sec. \ref{Seven}. 

\section{Basic concepts}
\label{One}
\subsection{Quantum optical circuit simulator}
In the context of this article a quantum optical circuit is the representation of a physical device made of linear optical elements where we can define photonic input and output states. The circuit
is defined by a transformation matrix \cite{KLM,Linear} that relates the bosonic creation operators at the input with the ones at the output. For example, if we consider a simple circuit with two input modes made of a single beamsplitter parametrized by two angles ($\theta$ and $\phi$), then the relationship between input and output operators is given as
\begin{equation}
	\begin{matrix}
		\hat{a}_1^\dagger \rightarrow & \cos(\theta)\hat{a}_1^\dagger + e^{-i \phi}\sin(\theta) \hat{a}_2^\dagger  \\
		\hat{a}_2^\dagger \rightarrow & -e^{i \phi} \sin(\theta)\hat{a}_1^\dagger + \cos(\theta) \hat{a}_2^\dagger.  
	\end{matrix}
\end{equation}
\newline
Alternatively, this same information can be expressed in a matrix representation which is more suited for coding:
\begin{equation}
	U=\begin{pmatrix}
	cos(\theta) & - e^{i \phi}sin(\theta)\\
	e^{-i \phi} sin(\theta) & cos(\theta)
	\end{pmatrix}\,.
\end{equation}

The output of this simple beamsplitter circuit with a single ket as input state is given by the following transformation:
\begin{equation}
\begin{split}
|n_1, n_2 \rangle &=\frac{\bigl(\hat{a}_{1}^{\dagger}\bigr)^{n_1}} {\sqrt{n_1!}}\frac{\bigl(\hat{a}_{2}^{\dagger}\bigr)^{n_2}}{\sqrt{n_2!}}|0, 0 \rangle \\
\rightarrow \frac{1}{\sqrt{n_1!n_2!}} &\Bigl( \hat{a}_{1}^{\dagger}\cos\theta +\hat{a}_{2}^{\dagger}e^{-i\phi}\sin\theta\Bigr)^{n_1} \\ 
& \Bigl(-\hat{a}_{1}^{\dagger}e^{i\phi}\sin\theta + \hat{a}_{2}^{\dagger}\cos\theta\Bigr)^{n_2}|0, 0 \rangle\,.
\end{split}
\end{equation}
where $n_1$ and $n_2$ are the number of photons in modes one and two at the input of the circuit. 

A quantum optical circuit simulator in its most basic form is a program 
that transforms an input state $\Psi_i$ of an optical circuit into an output state $\Psi_{o}$ using a matrix definition of the circuit. This matrix definition may be constructed
from the transformations associated with the individual optical circuit components and their connections. 

\subsection{SOQCS library}
The implementation of the calculation of partial distinguishability has been developed as part of the SOQCS library \cite{SOQCS}. SOQCS is a modular C++ library with a Python port which is aimed at obtaining the output states
and outcomes of optical quantum circuits under various imperfections. The configuration of the circuits is performed by enumerating their optical components and the connections between them. This library has a core simulator that is able to perform the calculation above for a basic general circuit given an arbitrary input state. SOQCS contains three different cores to calculate the quantum circuit output. 

In the direct core the output is obtained in a manner that is similar
to analytical calculations. This method scales poorly ($O(nn!)$ for each output ket where $n$ is the number of photons) but due to the simplicity of the operations it is faster than any other core
for a small number of photons (approximately $n \le 4$). Alternatively, each output ket amplitude may be obtained from a calculation of a matrix permanent that depends on the input state. Permanents are solved using
the Balasubramanian--Bax--Franklin--Glynn formula implemented in gray code \cite{Glynn}. This  method scales with a much better figure of $O(n2^n)$ for each output ket and it is the best one to solve
circuits with four or more photons.  These methods can be configured to obtain the full output distribution or to calculate only the amplitudes of a subset of output kets of interest.  

The computational cost of the calculation of the full output distribution is the cost of calculating a single amplitude by the number of possible output kets $O(n2^n \binom{n+d-1}{n})$ where $d$ is the number of degrees of freedom. Usually, the number of degrees of freedom is the number of channels $n_{ch}$ multiplied by the number of polarization modes $n_P$. However, if packet degrees of freedom are considered  then $d=n_{ch} \cdot n_P \cdot n_D$ where $n_D$
is the number of wavepackets in the simulation.

Additionally a third core for boson sampling calculation using Clifford A \cite{Sampling}  algorithm is also available. This method differs from the previous ones as it provides only
samples.  Clifford A algorithm has a cost by sample of order $O(d\cdot n\cdot 3^n)$ while its more efficient counterpart Clifford B has an improved cost of $O(n2^n+poly(n,d))$ by sample.
However, their limitation resides in the fact that only an approximation to the probability distribution can be obtained by means of repeated sampling.

Simulations using SOQCS can factor in the partial distinguishability between photons in an automated way based on their physical properties (emission time, shape, etc) using the implementation which is presented in this paper. This is based on an abstraction of the description of photon states that allows to configure the simulation in an intuitive way using a single instruction to enumerate the photons properties and the channel where they are initially set up. The library can provide both the output state of the circuit (this is the full set of the relevant probability amplitudes) as well as the measurement outcomes considering the effects of post-selection. The different outcomes can be expressed in terms of probability distributions or density matrices to account for mixed states which may result from some stochastic effects like noise or imperfections in the circuit elements. Post-selection conditions can be defined using detectors as virtual circuit elements. Furthermore, SOQCS considers a physical model of detectors that accounts for effects of efficiency, dead time and dark counts. A mechanism of calculation of losses is also implemented and the circuits can be initialized with photons
generated by a quantum dot using a suitable emission model in which effects of fine structure splitting, cross dephasing and spin scattering are considered.

\subsection{Definition of states}
In general, the states of an optical circuit are defined as a linear superposition of bosonic kets, each describing multiple occupation of several levels. That is,
\begin{equation}
|\Psi\rangle=\sum_i \alpha_i | n_{i,1}, n_{i,2}, ..., n_{i,n} \rangle\,,
\end{equation}
where $n_{i,j}\geq0$ is the occupation number of the level $j$ in the ket $i$.  A state can be represented in a computer as a list of kets where each ket consists of a 2-tuple made of a probability amplitude and a list of photon occupation numbers for each level,
\begin{equation}
\begin{split}
|\Psi\rangle&:= \{\{\alpha_1,\vec{v}_1\},\{\alpha_2,\vec{v}_2\}...\{\alpha_n,\vec{v}_n\}\}\,.
\end{split}
\end{equation}
Note that we are using an abstract definition of level in order to implement quantum states in computer simulations.
A level $j$ may be a combination of quantum numbers for the mode, polarization or any property that the photons may have in the considered device. 
The simulator is oblivious to the meaning of each level therefore it may be necessary to keep an index over the levels and their physical meaning. 
For example, $j\rightarrow\{ch_j,P_j\}$ where $ch$ is the mode number of level $j$ and $P_j$ is the polarization of the photons in that level.

\section{Basis: Mathematical formulation.}
\label{Two}
\subsection{Distinguishablity definition}
Using the definition of a photonic state presented above we are assuming that all photons are mutually indistinguishable. Using again the single beamsplitter circuit as a useful  example, we can 
see how an input with a photon in each channel is transformed when the beamsplitter is ideal and perfectly balanced ($\theta=\pi/2$ and $\phi=0$):
\begin{equation}
|1, 1 \rangle  \rightarrow -\frac{1}{\sqrt{2}}|2, 0 \rangle +\frac{1}{\sqrt{2}} |0, 2 \rangle 
\label{E9}
\end{equation}
In this case, a perfect bunching of photons in either one or the other output modes is found, and there is a zero probability of each photons taking different paths. This is the manifestation of the well known Hong-Ou-Mandel (HOM) effect \cite{HOM} in an ideal situation when photons are perfectly indistinguishable.

In real physical devices photons are rarely perfectly indistinguishable for a variety of reasons. They may have arrived at the beamsplitter at different times or with slightly different frequencies.  There are various ways to consider this distinguishability of the photons. In this article we are going to follow one approach that can be automated into a computer program.

A straightforward  way to include distinguishability is to extend the level definition with an extra label $j\rightarrow\{ch_j,P_j,D_j\}$ where $D_j$ refers to a wavepacket.
This label may represent, for example, two different fully distinguishable streams of photons. One stream arriving to the circuit at an early time $D=0$ and the other at a later time $D=1$. For the simple case of a single beamsplitter circuit we can consider an extended input state 
$|n_{ch=0,D=0},n_{ch=0,D=1},n_{ch=1,D=0},n_{ch=1,D=1}\rangle$
which leads to the output
\begin{equation}
\begin{split}
|1, 0, 0, 1 \rangle  \rightarrow &- 0.5\,| 1, 1, 0, 0 > - 0.5\,| 0, 1, 1, 0 > \\ 
&+ 0.5\,| 1, 0, 0, 1 > + 0.5\,| 0, 0, 1, 1 >\,,
\end{split}
\end{equation}
where all the possible outcomes have the same probability. This is the probability $0.25$ for both photons to be together in either channel 0 or 1, and the probability $0.5$ to be found in different channels. Alternatively, if we consider both streams to be indistinguishable (because, for example, they arrive at the same time) we recover the same output as in eq. \ref{E9} but represented by kets using the extended level definition,
\begin{equation}
|1, 0, 1, 0 \rangle  \rightarrow -\frac{1}{\sqrt{2}}|2, 0 , 0 , 0 \rangle +\frac{1}{\sqrt{2}} |0, 0, 2, 0 \rangle \,.
\end{equation}

\subsection{Partial distinguishability}
Two photons are identical if they share the same quantum numbers and spatial degrees of freedom (except the mode where they are traveling) otherwise they are different. 
Partial distinguishability between two photons can be defined as the partial overlap between the wavefunction of those photons. This partial overlap may occur for various reasons, most notably, small delays between photons as a consequence of their propagation in different paths.

Physically, the wavefunctions representing the spatial degrees of freedom of the photons are an infinite set of continuous functions.
To implement partial indistinguishability in a simulator we restrict the photons to exist in a discrete finite subset of those wavefunctions. We call the elements of this subset wavepackets and they are characterized by their central times and frequencies $|P_i\rangle=|\Psi_{t_{0_i},\omega_{0_i}}\rangle$. We are making no assumption at this moment about the particular shape of these wavepackets.

The spatial wavefunctions are therefore treated as an additional discrete quantum number like in the previous section. The only issue that remains is that now our base is non-orthogonal $\langle  P_i | P_j \rangle\neq 0$. This can be solved using the Gram-Schmidt orthonormalization procedure as suggested in ref. \cite{Distin},
\begin{equation}
\begin{split}
|\tilde P_0 \rangle &= | P_0 \rangle ,  \\
|\tilde P_1 \rangle &= \frac{ | P_1\rangle -  | \tilde P_0 \rangle \langle \tilde P_0 | P_1 \rangle } {\sqrt{1-|\langle \tilde P_0 | P_1 \rangle|^2 }} , \\
|\tilde P_2 \rangle &= \frac{ | P_2 \rangle-  | \tilde P_0 \rangle \langle \tilde P_0 | P_2 \rangle - | \tilde P_1 \rangle \langle \tilde  P_1 | P_2 \rangle   } {\sqrt{1-|\langle \tilde P_0 | P_2 \rangle |^2- | \langle \tilde P_1 | P_2 \rangle|^2 }} \,. 
\end{split}
\end{equation}

The rules of transformation of the input wavepackets to an orthonormal basis are obtained from the result above,
\begin{equation}
\begin{split}
| P_0 \rangle &\rightarrow | \tilde P_0 \rangle c_{0,0} \\
| P_1 \rangle &\rightarrow | \tilde P_0 \rangle c_{1,0} + |\tilde P_1 \rangle   c_{1,1}  \\
| P_2 \rangle &\rightarrow | \tilde P_0 \rangle c_{2,0} + |\tilde P_1 \rangle   c_{2,1}   + | \tilde P_2 \rangle c_{2,2} \,.
\end{split}
\label{ETimes}
\end{equation}
where $c_{i,j}$ are numerical coefficients.

This transformation, which can be expressed as a matrix that relates the non-orthogonal wavepackets with their orthogonal counterparts, can now be integrated with the matrix representing the original circuit, where photons sharing the same quantum numbers are assumed indistinguishable. The result, obtained by mere matrix multiplication, is a matrix representation of the entire circuit including the fact that photons are partially distinguishable.

The transformation matrix for partially distinguishable photons plays the role equivalent to the definition of the emitter in the simulation. Moreover, it is used in the same way as a matrix representing any other optical element of the circuit. As shown in the previous section, the results obtained for photons defined as Fock states are correctly reproduced provided these are replaced by orthogonal wavepackets. 
This transformation, based on Gram-Schmidt orthonormalization, allows to fulfill this condition. Now the spatial degrees of freedom can be treated as a discrete index of fully orthornormal wavepackets.
Note that due to the non-unitarity of this matrix, initial states cannot contain superpositions of the original non-orthogonal wavepackets but each set of photons has to be initialized to a definite wavepacket.

\subsection{Delays}
The previous subsection explains how to incorporate partially distinguishable photons at the input stage of a circuit. Nevertheless, delays between photons may be engineered into a circuit, for example, a Mach-Zehnder interferometer in which it will affect the relative overlap between the photons.

To introduce a delay $\Delta t$ in the circuit, we consider the temporal scale of the simulation to be divided into periods of length $\Delta t$.  A delay causes photons defined in one period to be moved into the next one. Photon wavepackets within the same period can represent partially distinguishable photons. On the other hand, wavepackets at different periods do not overlap as we assume long delay times with respect the width of the packets. Therefore, a delay will imply a creation of a new group of packets which are equal to the previous one but delayed one period to account for the extra possible spatial wavefunctions that a photon may take,
\begin{equation}
\begin{split}
| t_0 \rangle \rightarrow &| \tilde t_0 \rangle c_{0,0} \\
| t_1 \rangle \rightarrow &| \tilde t_0 \rangle c_{1,0} + |\tilde t_1 \rangle   c_{1,1}  \\
| t_2 \rangle \rightarrow &| \tilde t_0 \rangle c_{2,0} + |\tilde t_1 \rangle   c_{2,1}   + | \tilde t_2 \rangle c_{2,2} \\
| t_3 = t_0 + \Delta t \rangle \rightarrow & 
| \tilde t_3 \rangle c_{0,0}  \\
| t_4 = t_1 + \Delta t\rangle  \rightarrow & 
| \tilde t_3 \rangle c_{1,0} + | \tilde t_4 \rangle c_{1,1}  \\
| t_5 = t_2 + \Delta t\rangle  \rightarrow & 
| \tilde t_3 \rangle c_{2,0} + | \tilde t_4 \rangle c_{2,1} + | \tilde t_4 \rangle c_{2,2}   \,,
\label{FGram}
\end{split}
\end{equation}
where we change the notation to define wavepackets of central time $t$ as $|t_i\rangle$.
Alternatively, photons may be created at different periods and brought together to the same period by the delay. The overlaps of the packets between different periods are zero while the coefficients between packets within the same period are the same as in the previous case because we consider all wavepackets to be delayed by the same amount of time. For this reason, the Gram-Schmidt coefficients are the same for the wavepackets in different periods and, more importantly, the orthonormal components of the packets at different periods remain constant but delayed in time. Therefore the delay operation can be written as, 
\begin{equation}
\begin{split}
\hat{T}_D =\sum_{j=0}^{n_t} | \tilde t_{j+n_t} \rangle  \langle \tilde t_j |
\end{split}
\end{equation}
where $n_t$ is the number of packets in a period.

This operation written in a matrix form can be treated on the same footing as the beamsplitter matrix or any other circuit element matrix. The whole circuit matrix can be built multiplying the single element matrices in their order of operation. Note that this is not an unitary operation like that of a beasmplitter because the delay of a wavepacket is in part a classical operation. The strategy of discretizing the wavepacket space and rewriting the operation as a matrix works because optical circuits are interpreted as one way operations. This is, we have a clearly defined transformation from input to output and hence the lack of reversibility in the operation is not a problem.

\subsection{Gate implementation}
To implement these operations we can define an object circuit that contains the index between levels and their physical meaning and the circuit matrix. Each time we add
a linear circuit element (like a beamsplitter) this circuit matrix is updated. The initial configuration of the photon wavepackets and the delay may be treated in the same manner as other linear optical elements,
despite the fact that their corresponding matrices are not unitary as explained above. This results in two operations: one for the emitter configuration 
\begin{verbatim}
void emitter(matc G);
\end{verbatim}
where \texttt{G} is the Gram-Schmidt transformation matrix, and one to introduce a delay of one period in the circuit
\begin{verbatim}
int delay(int i_ch);
\end{verbatim}
where \texttt{i\_ch} is the channel where the delay takes place.

\section{Level 0: Overlap matrix}
\label{Three}
The inconvenience of the previous operations is that they are very cumbersome to use directly because they require a Gram-Schmidt orthonormalization to be performed in advance. The Gram-Schmidt  coefficients $c_{i,j}$ can be obtained automatically from the hermitian matrix $S$ where $S_{i,j}=\langle P_i | P_j \rangle$. For this purpose we employ a Cholesky decomposition \cite{Cholesky} for which various methods are available in standard matrix libraries \cite{Eigen3}. 

A Cholesky decomposition can be performed only if the overlap matrix $S$ is positive definite. If this is not the case but the negative eigenvalues are few and small then approximative methods are available  \cite{ModCholesky}. These methods are called modified Cholesky decomposition methods.

The methods for the modified Cholesky decomposition rely on adding to the original overlap matrix $S$ a correction matrix $\Delta E $ where the values of its elements are small $\tilde S = S + \Delta E$. Different methods use a slightly different approaches to obtain this $\Delta E$. However, these methods rely on pivoting 
to minimize the error introduced by this correction. These algorithms perform well maintaining the error small, however pivoting alters the meaning of the decomposition. This is usually not an issue in computing Newton-like gradients for which these methods are usually used.

Therefore, we use a custom method that uses no pivoting to perform a modified Cholesky decomposition when needed. First, the eigenvalue matrix $D=U^\dagger S U$ is obtained. Then the small negative
eigenvalues are updated with positive ones $\tilde D_i=min(|D_i|,\epsilon)$ where $\epsilon$ is a small value and the corrected overlap matrix is reconstructed $\tilde S=U\tilde D U^\dagger$. Finally,
a Cholesky decomposition is performed for the positive definite matrix  $\tilde S$. This very simple method has the inconvenience that the error is unbounded and therefore a check of the solution has 
to be performed afterwards. This check is carried out by computing the normalization of the rows of the resulting matrix after performing the decomposition. Usually, the magnitude of the error can be minimized by carefully selecting the leading wavepacket.

We modify the operation, described in the previous section, in the circuit object, so it uses an overlap matrix instead of Gram-Schmidt coefficients, as follows 
\begin{verbatim}
void emitter(matc S);
\end{verbatim}
where \texttt{S} is the overlap matrix; the delay operation can be used unchanged. These operations are implemented in ref. \cite{SOQCS} as the two most basic ways to control photon distinguishability.

\section{Levels 1: The photon shape model.}
\label{Four}
\subsection{Coefficient calculation.}
To calculate the hermitian overlap matrix $S$, we have assumed that we know the overlaps between all pairs of wavefunctions, and that the wavefunctions can be modeled as wavepackets.
These wavepackets are centered around a mean frequency and are defined
as the integral in time of a phase and an envelope for each one of the possible times where those wavefunctions have a non-zero probability to exist, 
\begin{equation}
| P_i \rangle =  \int dt_i 	K(t_i) e^{-i \omega_i (t_i-t_{i_0})} | t_i \rangle \,,
\end{equation}
where the envelope function $K$ is a real function that is zero at $\pm \infty$.

For example, we may assume that the envelope has a Gaussian shape centered around a particular emission time $t_{i_0}$ and frequency $w_{i_0}$,
\begin{equation}
| P_i \rangle =  \int dt_i 	\left( \frac{\sqrt{\Delta \omega_i}}{\pi^{1/4} } \right) e^{-(t_i-t_{i_0})^2 \Delta \omega_i^2 } e^{-i \omega_{i_0} (t_i-t_{i_0})} | t_i \rangle \,,
\end{equation}
where $\Delta \omega_i$ is the width of the Gaussian wavepacket. Then the overlap between two photons becomes,
\begin{equation}
S_{i,j}= \langle P_i | P_j \rangle =N e^{-T(t_{i_0}-t_{j_0})^2} e^{-W (\omega_{i_0}-\omega_{j_0})^2} e^{-i \phi (t_{0i}-t_{0j})} \,.
\end{equation}
with an exponential term that depends on the relative position of the wavepackets in time, another term that depends on their central frequency difference and a phase term. The coefficients in the exponents are
\begin{equation}
T=\frac{1}{2}\frac{ {\Delta \omega_i}^2 {\Delta \omega_j}^2}{ {\Delta\omega_i}^2+{\Delta\omega_j}^2}\,,
\end{equation}
\begin{equation}
W=\frac{1}{2}\frac{1}{{\Delta\omega_i}^2+{\Delta\omega_j}^2 }\,,
\end{equation}
and 
\begin{equation}
\phi=\frac{ {\Delta \omega_i}^2 {\omega_{j_0}}+{\Delta \omega_j}^2  {\omega_{i_0}}}{{\Delta\omega_i}^2+{\Delta\omega_j}^2 }\,.
\end{equation}
The normalization constant is found to be,
\begin{equation}
N=\sqrt{2} \frac{\sqrt{\Delta\omega_i \Delta\omega_j} }{\sqrt{{\Delta\omega_i}^2+{\Delta\omega_j}^2}}\,.
\end{equation}

In the particular case where the two Gaussian shaped wavefunctions are equal ($\omega_{i_0}=\omega_{j_0}=\omega$ and $\Delta \omega_i=\Delta \omega_j=\Delta \omega$)  but delayed in time with respect to each other we recover the result of ref. \cite{Distin},
\begin{equation}
\langle  P_i | P_j \rangle = e^{i (t_j-t_i) \omega} e^{-\frac{(t_i-t_j)^2}{4} \Delta \omega^2 } \,,
\end{equation}
which is used to calculate the outcome probabilities for various groups of photons arriving to a beamsplitter. 
These probabilities are also reproduced numerically in our examples as a validation of the implementation presented in this paper.

\subsection{Table definition.}
The wavepacket shape model is a table stored for convenience as part of the circuit object that contains wavepacket definitions. Wavepacket definitions can be straightforwardly summarized in a single table with one column for each wavepacket where each parameter is stored in a different row. These rows are, wavepacket index, wavepacket central time, wavepacket central frequency and width of the wavepacket (when the packet shape is Gaussian). For other wavepacket shapes, like an exponential, the second parameter may be the exponential characteristic decay time.  

As a consequence, the emitter operation definition is simplified to
\begin{verbatim}
veci emitter ( int npack, matd packets);
\end{verbatim}
where \texttt{packets} is the table with the wavepacket definitions and \texttt{npack} the number of wavepackets in the table. The wavepackets shape is established while creating the circuit.

\section{Levels 2 and 3: Increasing the abstraction}
\label{Five}
In general, there are no restrictions in the way that the table with the wavepacket definitions can be created. Each column can be added at different times when suitable.
Instead of creating a big table with all the wavepacket definitions at the beginning of a simulation it is more convenient to add the columns whenever they are needed to define a new wavepacket.
The code that describes the shape of the wavepacket is

\begin{verbatim}
int def_packet( int n, double t, double f, 
                double w);
\end{verbatim}
where \texttt{n} is the suggested packet number, \texttt{t} is the time, \texttt{f}  the frequency and \texttt{w} is a variable controlling the width of the wavepacket according to a predetermined shape. If various periods are declared to handle delays this instruction will create the multiple packets needed in each period. An additional check is present to avoid duplicated entries of the table.

The independent definition of wavepackets allows for one additional abstraction. Photons can be treated as entities that can be related to the circuits.
A set of input photons consists of an input state and a table defining the wavepackets. Each time a group of photons is created the occupation of the initial
state is updated and a new photon wavepacket configuration is stored if necessary. This is done with,

\begin{verbatim}
int add_photons( int N, int ch, int P, 
				  double t, double f,double w);
\end{verbatim}

where \texttt{N} is the number of photons to be created, \texttt{ch}  and  \texttt{P} the mode and polarization of the photons while \texttt{t}, \texttt{f} and \texttt{w} have
the same meaning as in \texttt{def\_packet}. Once the photons are created the instruction,
\begin{verbatim}
void send2circuit();
\end{verbatim}

calls the circuit \texttt{emitter} to configure the wavepackets and calculate the Gram-Schmidt coefficients. This last instruction is handled internally and executed when the user defines the last detector.
A larger \verb|qodevice| class contains both the definition of the optical circuit and its initial state for increased transparency. The advantage of this abstraction is that photons can be created independently whenever it is convenient or required. This is carried out by providing the parameters that define the wavepackets' discrete and continuous degrees of freedom. 
Their distinguishability properties are then calculated and applied automatically.

\section{Examples}
\label{Six}
\subsection{HOM Visibility}
\begin{figure}[h]
  \centering
  \includegraphics[width=0.5\textwidth]{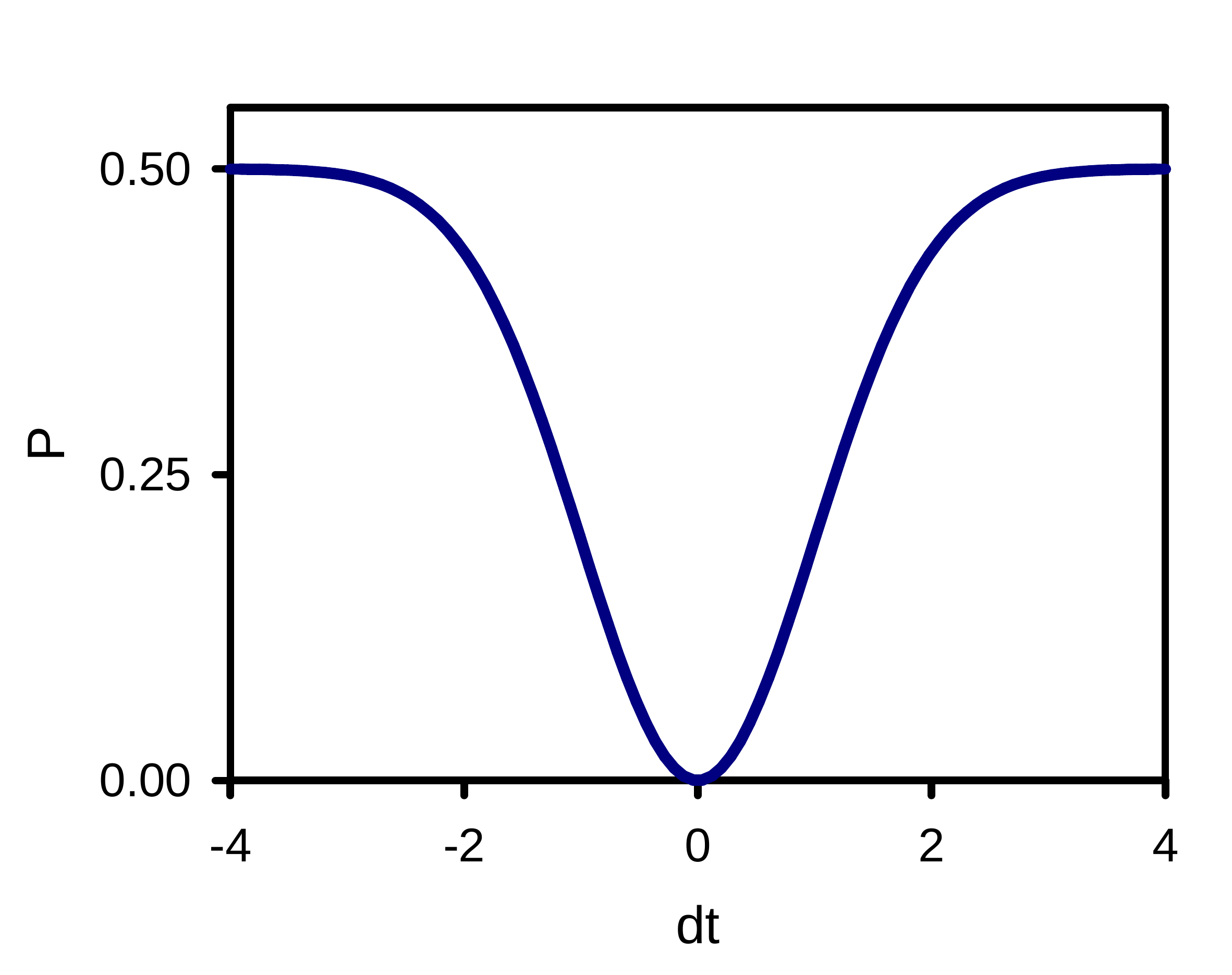}
  \caption{Numerical calculations of HOM visibility.}
  \label{F2}
\end{figure}

The most basic example of photon partial distinguishability itself is the measurement of HOM visibility. With that purpose we consider a circuit
with two photons of Gaussian shape arriving in different modes to a balanced beamsplitter with a relative delay \texttt{dt} between them. The calculation
of the probability of these photons to be in two different modes at the output is performed using the SOQCS library \cite{SOQCS} where the distinguishability model presented in this paper has been
implemented,
\begin{verbatim}
example->add_photons(1,0, H, 0.0, 
                     1.0, 1.0);
example->add_photons(1,1, H,  dt, 
                     1.0, 1.0);
example->beamsplitter(0,1,45.0,0.0);
example->detector(0);
example->detector(1);
\end{verbatim}
Note that here the process of sending the photons to the circuit is done internally and automatically as part of the circuit definition upon detection that the circuit is completed (this is, the last detector is declared). The result of the calculation (in Fig. \ref{F2}) shows the characteristic dip of the HOM effect. The probability for two perfectly indistinguishable photons to leave the beamsplitter in two different modes is zero while if they are distinguishable each photon has the same probability to end in any of the two outgoing modes. The delay between photons makes them distinguishable or indistinguishable depending on the overlap between the Gaussian wavepackets.

\begin{figure}[h]
  \centering
  \includegraphics[width=0.5\textwidth]{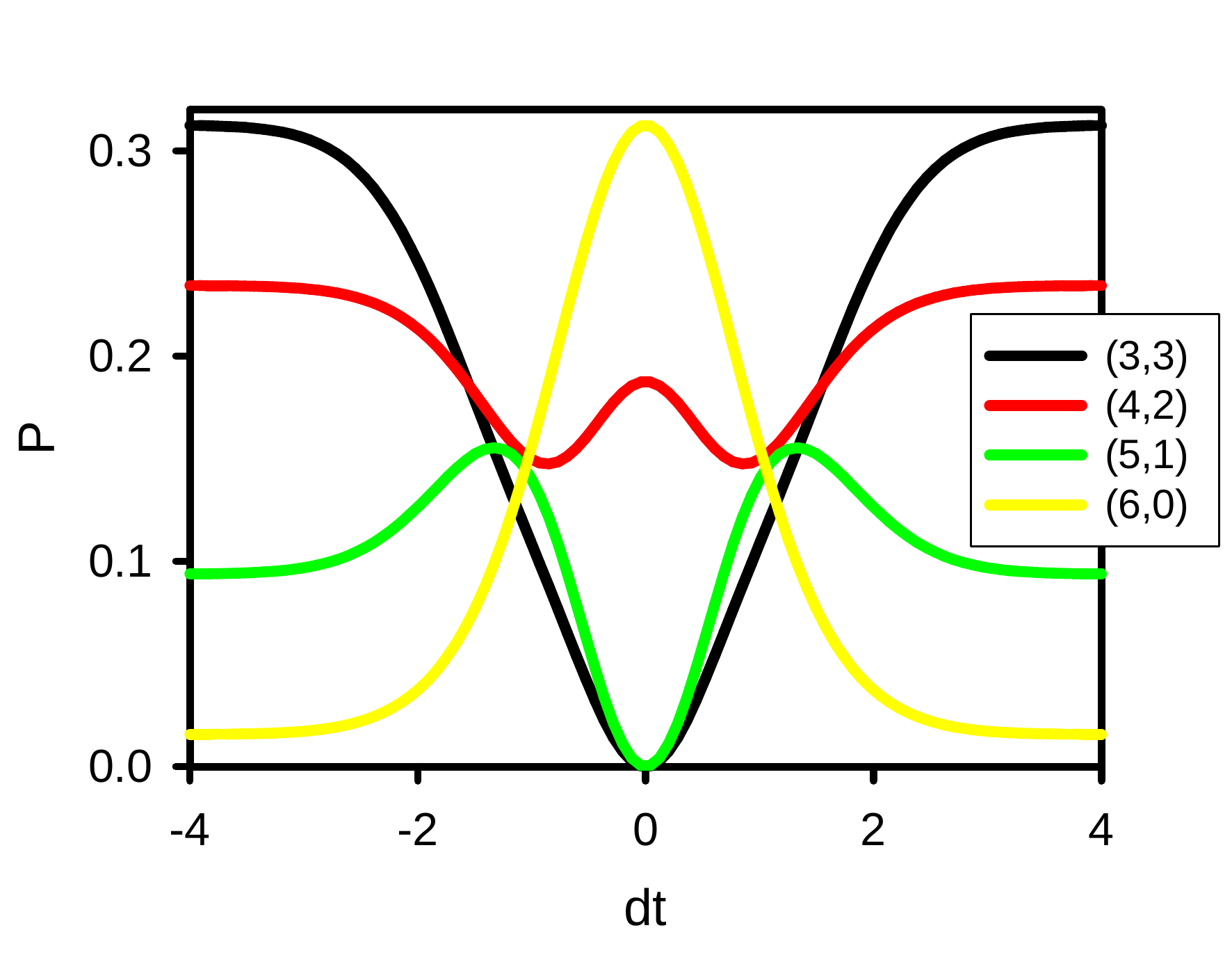}
  \caption{ Probability of the different outcomes for the case when two groups of three photons each arrive to a beamsplitter with a delay dt. Our results, which uses the automated code presented in this paper and implemented in SOQCS library \cite{SOQCS}, are in agreement with the results presented in ref. \cite{Distin} (Fig. 12a).}
  \label{F3}
\end{figure}

This calculation can be modified to initialize the circuit with three photons in each channel. In this case we reproduce the results presented in ref.\cite{Distin} where distinguishability is calculated in a similar manner (see Fig. \ref{F3}) . This example is a good illustration of a case where various photons share the same wavepacket. See how constructive interference increases the probability of some of the outcomes for moderate values of the delay in a way that can not be obtained from classical calculations.

\subsection{Delay gate and correlation}
\begin{figure}[h]
  \centering
  \includegraphics[width=0.5\textwidth]{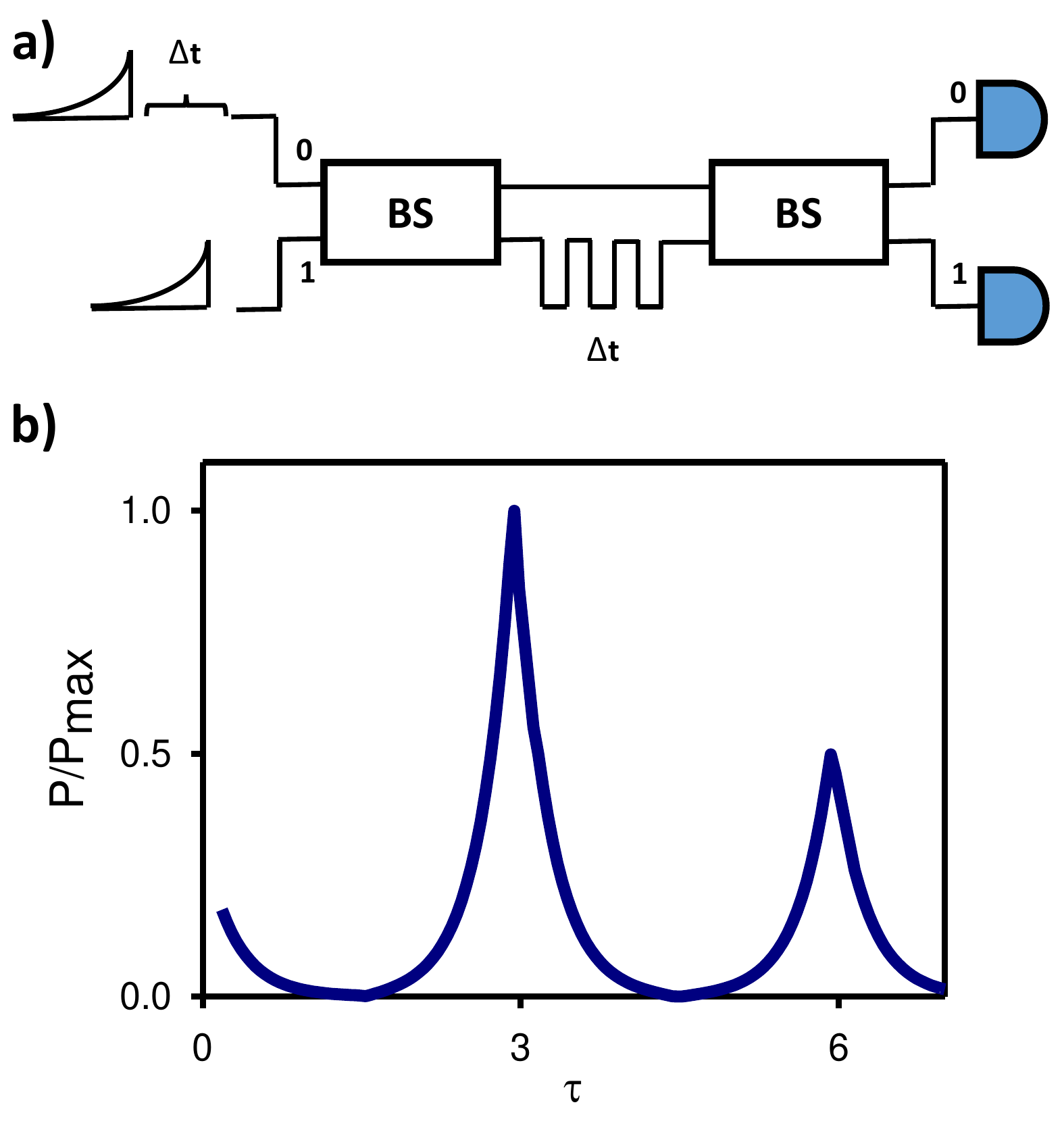}
  \caption{ a) Schematic of the simulated circuit. Two photons of exponetial shaped wavepackets arrive to the circuit with a delay dt. The circuit consists in two balanced beamsplitters 
  with a delay gate matching the photon delay between the two beamsplitters. b) Probability of two photons to arrive with a time difference $\tau$ at the modes 0 and 1 respectively.}
  \label{F4}
\end{figure}

The simplest circuit that requires an explicit delay operation is the one presented in fig.\ref{F4}a. We consider a circuit with two interconnected beamsplitters with a delay in one of the modes between the two
beamsplitters matching the input photon delay,

\begin{verbatim}
example->add_photons(0,0, H,    t2, 
                        1.0,  0.01);
example->add_photons(0,1, H,    t1, 
                        1.0,  0.01);
example->add_photons(1,0, H,  0.001,
                         1.0,  0.3);
example->add_photons(1,1, H,  3.101, 
                         1.0,  0.3);
example->beamsplitter(0,1,45.0,0.0);
example->delay(1);
example->beamsplitter(0,1,45.0,0.0);
example->detector(0);
example->detector(1);
\end{verbatim}

In this case we calculate the probability of two photons arriving at two different modes with a time difference $\tau=t_1-t_2>0$. This is an example where wavepackets are defined such that no photon is to be found in the input. The instruction \verb|example->add_photons(0,1, H, t1, 1.0, 0.01);| adds zero photons to the input but creates a wavepacket that photons may have at the output. Note that in this case the two extra measurement packets are made narrow to avoid overlap between them while performing the sweep of the parameters $t_1$ and $t_2$. It is also important to remember
that the times are always relative to the leading wavepacket. The output probability is plotted in fig. \ref{F4}b where we can see the characteristic peaks of this kind of experiments at $\Delta t$ and $2\Delta t$.

This model is a representative of measurement protocols where a Mach-Zehnder interferometer is used to introduce a delay \cite{Delay}. An intensity correlation is measured at the output. Two peaks appear in Fig. \ref{F4}b, at three and six time units due the different paths that a pair of photons can take in the circuit. We also consider partial distinguishability between photons; this implies that a small peak is also found at zero time.

\subsection{Entanglement swapping circuit}
\begin{figure}[h]
  \centering
  \includegraphics[width=0.45\textwidth]{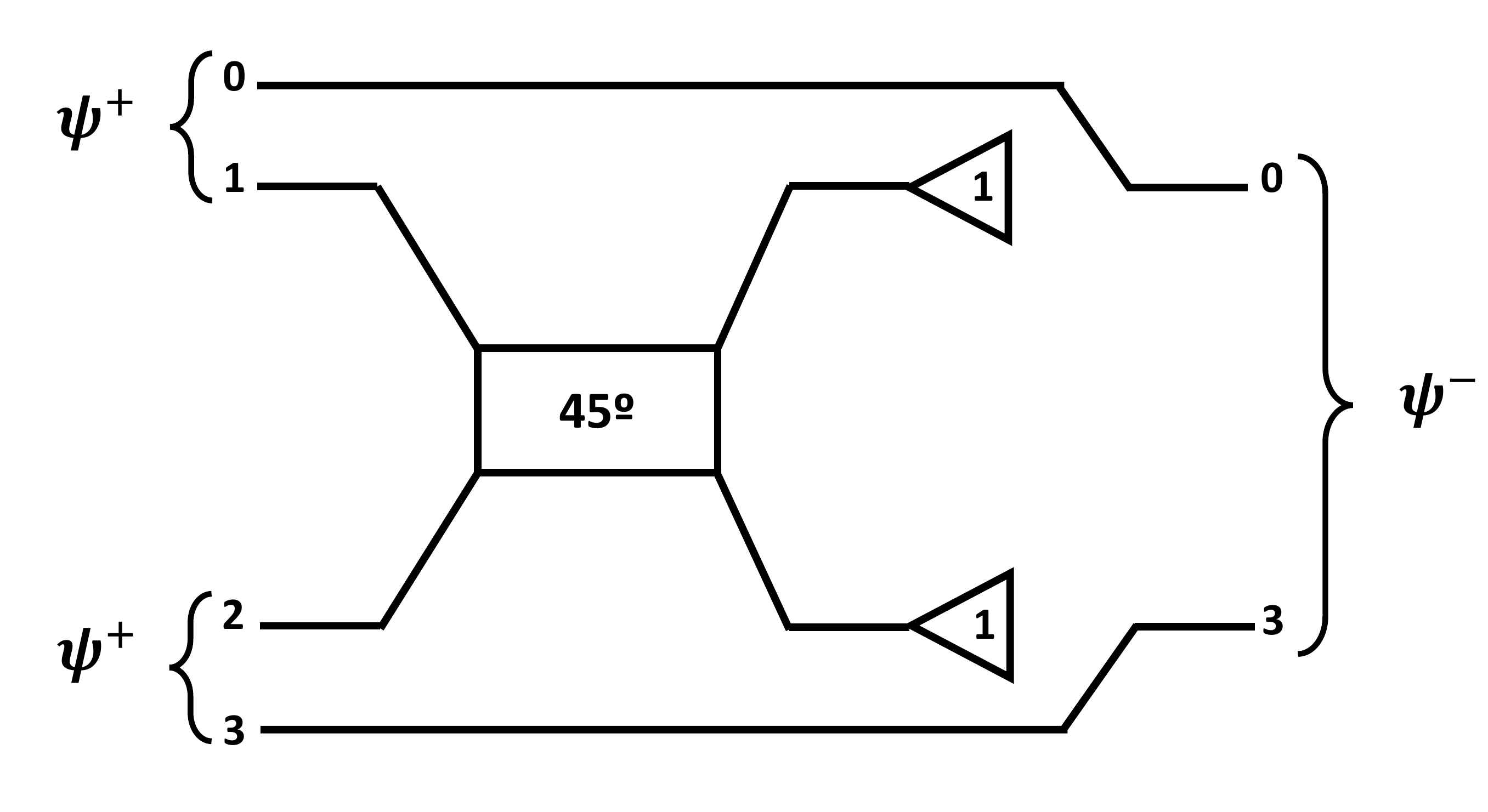}
  \caption{Schematic of an entanglement swapping protocol circuit as presented in ref. \cite{Entanglement}. The protocol is performed in a four channel circuit with one photon in each channel. Each pair of photons are entangled to each other. At the output if the post-selection condition in channels 1 and 2 is met the two previously unentangled photons in channels 0 and 3 become entangled.}
  \label{F5}
\end{figure}

In Fig. \ref{F5}, the optical circuit implementation of an entanglement swapping protocol \cite{Entanglement} is shown. 
Below the code written in SOQCS can be found which allows a simulation of this circuit for a photonic input state $|\Phi^+\rangle_{0,1}=\frac{1}{\sqrt{2}}(|HH\rangle_{0,1}+|VV\rangle_{0,1})$ in channels 0 and 1 and the same state for the photons in channels 2 and 3. 
Thus the total input is $|\Psi\rangle_{0,1,2,3}=|\Phi^+\rangle_{0,1,2,3}\otimes|\Phi^+\rangle_{0,1,2,3}$. 
For a set of indistinguishable photons the output corresponding to this input would be $|\Psi^-\rangle_{0,3}=\frac{1}{\sqrt{2}}(|HV\rangle_{0,3}-|VH\rangle_{0,3})$ that leads to the output density matrix,
\begin{scriptsize}
\begin{verbatim}
 | H(0)0, V(0)3 >  0.5000 -0.5000 
 | V(0)0, H(0)3 > -0.5000  0.5000 
\end{verbatim}
\end{scriptsize}
Below we simulate the existence of a relative delay between the photons interacting in the beamsplitter. The overlap between those two photons is reduced to a value equal to \texttt{0.6065} due this delay.
\begin{verbatim}
# Build the circuit
eswap->add_BellP(0, 1, 'p', 0.0,
                  0.0, 1.0, 1.0,
                 10.5, 1.0, 1.0)
eswap->add_BellP(2, 3, 'p', 0.0, 
                 0.01, 1.0, 1.0, 
                 10.0, 1.0, 1.0)
eswap->beamsplitter(1,2,45.0,0.0)
eswap->detector(0)
eswap->detector(1,1)
eswap->detector(2,1)
eswap->detector(3)
\end{verbatim}
The corresponding density matrix is obtained, 
\begin{scriptsize}
\begin{verbatim}
 | H(0)0, H(2)3 >  0.1412  0.0000  0.0000  0.0000 
 | H(0)0, V(2)3 >  0.0000  0.3588 -0.2176  0.0000 
 | V(0)0, H(2)3 >  0.0000 -0.2176  0.3588  0.0000 
 | V(0)0, V(2)3 >  0.0000  0.0000  0.0000  0.1412 
\end{verbatim}
\end{scriptsize}
which gives the same result as the analytic formulation given in ref.\cite{Entanglement} for the same value of the overlap. We can see how the partial distinguishability between photons
leads to the loss of purity of the density matrix.

\section{Conclusions}
\label{Seven}
We have presented an implementation of partial photon distinguishability, based on a Gram-Schmidt orthonormalization, in a computer simulation. It consists of treating the continuous degrees of freedom of the photons
as instances of a set of discrete non-orthonormal wavepackets. The overlap between the wavepackets is used to create an orthonormal basis in which it is possible to carry out the calculation treating the wavepacket number in the same footing as a quantum degree of freedom. This also allows to use delays in the same manner as the rest of the linear optical elements even if in a strict way delays are not unitary operations. The output state of the circuit can be processed to provide physically meaningful information about their probability distributions and amplitudes.

This implementation can be found as part of the simulation package SOQCS \cite{SOQCS}. The main advantages of this implementation are two fold. In one hand it can be configured in an automatic manner from physical parameters. On the other hand, this implementation is modular and independent  of other effects like losses or the simulation method. This is important in the context of a larger library because it allows for integration with different kinds of simulations without a need of any fundamental library reformulation.

\section*{Acknowledgements}
This work has received funding from the Enterprise Ireland’s DTIF programme of the Department of Business, Enterprise and Innovation, project QCoIr Quantum Computing in Ireland: A Software Platform for Multiple Qubit Technologies No. DT 2019 0090B. We acknowledge discussions and support from Tyndall National Institute and Rockley Photonics Ltd. We also thank to Paul Watts for useful comments on the manuscript.

\bibliographystyle{elsarticle-num} 
\bibliography{A1Refs}

\end{document}